\documentclass[letter]{aa}

\usepackage{graphics}
\usepackage{epsfig}

\begin{document}

\title{Microlensing of Circumstellar Envelopes}
\subtitle{III.  Line profiles from stellar winds in homologous
expansion}
\titlerunning{Microlensing of stellar winds in homologous expansion}
\author{M.~A.~Hendry \inst{1,2}
  \and R.~Ignace \inst{2,3}
  \and H.~M.~Bryce \inst{1,3}
   }
\authorrunning{M.~A.~Hendry et al.}
\offprints{M.A..~Hendry}

  \institute{Department of Physics and Astronomy, University of
  Glasgow, G12 8QQ, UK
        \and
  Department of Physics, Astronomy and Geology, Box 70652,
              East Tennessee State University, Johnson City, Tennessee,
              37614, USA
         \and
             Department of Astronomy, University of Wisconsin,
             475 North Charter Street, Madison, Wisconsin, 53706, USA
             }

\date{Received ...; Accepted ...}

\abstract{ This paper examines line profile evolution due to the
linear expansion of circumstellar material obsverved during a
microlensing event. This work extends our previous papers on
emission line profile evolution from radial and azimuthal flow
during point mass lens events and fold caustic crossings. Both
``flavours'' of microlensing were shown to provide effective
diagnostics of bulk motion in circumstellar envelopes. In this work
a different genre of flow is studied, namely linear homologous
expansion, for both point mass lenses and fold caustic crossings.
Linear expansion is of particular relevance to the effects of
microlensing on supernovae at cosmological distances. We derive line
profiles and equivalent widths for the illustrative cases of pure
resonance and pure recombination lines, modelled under the Sobolev
approximation. The efficacy of microlensing as a diagnostic probe of
the stellar environs is demonstrated and discussed.

}

\maketitle

\section{Introduction}

Recently gravitational microlensing has been shown to be a powerful
tool for probing not only the nature and distribution of dark matter
but also the astrophysical properties of the source being lensed
(see, for example, Gould 2001 for a comprehensive review). While the
vast majority of microlensing events observed to date can be
adequately approximated as point sources lensed by a point mass
lens, there is a small but significant fraction of events in which
the source must be modelled as {\em extended\/}, which has important
consequences for the evolution of the broad-band lightcurve and
spectrum during the microlensing event and produces observational
signatures that are now readily detectable -- chiefly due to the
introduction of microlensing `alert' networks. This development has
allowed intensive photometric and spectroscopic monitoring of many
microlensing events, spearheaded by the PLANET collaboration (Albrow
et al. 1998).  The alert strategy is especially well-suited to fold
caustic crossings -- in e.g. close binary lensing events -- since
observation of the first caustic crossing permits prediction of the
second caustic crossing, allowing intensive follow-up observations
to be scheduled. Moreover, every fold caustic crossing must be
treated as an extended source event, making them ideal probes of the
astrophysics of the source.

\noindent In the past few years `alert response' broad-band
photometric observations of extended source microlensing events have
shown clear detections of stellar limb darkening on the photospheres
of a number of stars, including: K giants (Albrow et al. 1999;
Albrow et al. 2000; Fields et al 2003), a metal-poor A dwarf in the
Small Magellanic Cloud (Afonso et al. 2000), a G/K subgiant (Albrow
et al. 2001a) and an F8-G2 solar-type star (Abe et al. 2003). These
observations have been used to estimate limb darkening parameters
and to constrain stellar atmosphere models, comparing for example
the LTE ATLAS models of Kurucz (1994) with the PHOENIX `Next
Generation' models of Hauschildt et al. (1999).

\noindent In addition Albrow et al. (2001b) used the VLT to carry
out observations of $H\alpha$ equivalent width variation in the
atmosphere a K3 bulge giant during a fold caustic crossing event.
These authors showed that highly resolved observations of the
microlensed spectrum could be a powerful discriminant between
different stellar atmosphere models, and indeed could also be a
useful method for breaking the near-degeneracy between the
parameters of different lens models. This work essentially
vindicated the theoretical modelling of Heyrovsk\'{y}, Sasselov and
Loeb (2000), which predicted that with 8m-class telescopes it should
be possible to use highly time- and spectrally-resolved observations
of microlensing events to carry out `tomography' of stellar
atmospheres.

\noindent A notable feature of the theoretical literature on
extended source microlensing events, however, has been the
comparative neglect of extended {\it circumstellar\/} envelopes.
Although the model developed in Coleman~(\cite{coleman1998}) and
Simmons et al. (2002) includes such a scattering envelope, and
Coleman et al. (1997) also considered the case of a non-spherical
envelope (e.g., a Be~disk), these computations were carried out only
for the broad-band photometric and polarimetric response to a
microlensing event.

\noindent In two preceding papers, Ignace \& Hendry (1999) and
Bryce, Ignace and Hendry (2003) calculated the microlensing
signatures of emission line profiles from expanding and rotating
spherical shells, lensed by a point mass lens and a fold caustic
respectively. Their results were for highly simplistic velocity
fields -- constant expansion or constant rotation -- in order to
isolate the effects of the microlensing and demonstrate clearly the
diagnostic potential for deriving velocity information about the
shells.  The authors showed that, while in the absence of lensing
the integrated line profiles for expanding and rotating shells
yielded identical `flat top' profiles in each case, microlensing
clearly breaks the degeneracy between these cases.

\noindent In this third paper, we choose to focus on the particular
case of circumstellar media in homologous expansion, with radial
velocity $v(r) \propto r$.  This case is interesting for two
reasons.  Firstly, this type of velocity law applies to the early
phases of nova and supernova (SN) explosions.  Since the latter are
observed to cosmological distances, it becomes increasingly likely
that such events will be lensed as the light from the SN makes its
way from a distant galaxy to the Earth (c.f. Dalal et al. 2003).
Thus a consideration of homologous expansion is relevant if only to
investigate definite signatures of microlensing from emission
profile shapes observed in SN events.  Indeed Bagherpour et al.
(2004) consider the microlensed lightcurves of Type Ia SN, and
Bagherpour et al. (2005) extend their analysis to consider the
impact of microlensing on P Cygni profiles in the spectra of Type Ia
SN.

\noindent Secondly, in this paper we employ standard Sobolev theory
to model the emission line profiles; the Sobolev approximation is
valid for the highly supersonic flows commonly found in SNe and
stellar winds. Furthermore the case of homologous expansion leads to
a significant simplification of the expressions and a consequent
substantial reduction in computational effort.  Thus it is
worthwhile studying this case to better appreciate the
characteristics of how microlensing will affect emission lines
formed in winds with more general velocity laws.

\noindent The structure of this paper is as follows.  In Section
\ref{sec:lft} we describe the basic features of line formation
theory, and the Sobolev approximation which we employ. In Section 3
we derive, and briefly discuss, unlensed profiles for the case of
pure resonance and pure recombination lines, in a stellar wind
undergoing homogolous expansion. In Section 4 we present
illustrative results, calculating the time evolution of line profile
and equivalent width for pure resonance and pure recombination lines
lensed by a point mass lens and a fold caustic. Finally in Section 5
we discuss our results and their future applicability in a number of
specific astrophysical contexts.

\section{Line formation theory}
\label{sec:lft}

\begin{figure}
\centerline{\epsfig{file=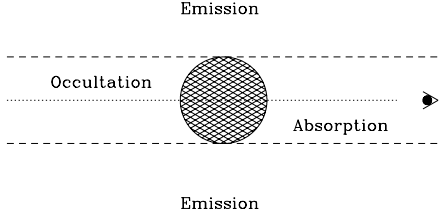,angle=0,width=7cm}}
\caption{\small{A schematic showing the contribution from different
regions to the line profile shape, for an observer on the right hand
side. Much of the line emission is produced around the circular tube
indicated in projection by the dashed lines. On the far side (left
hand side) of the tube the emission will be occulted by the star. On
the near side, the circumstellar material will attenuate the
continuum emission from the central photosphere.}}
\label{wind_schematic}
\end{figure}

\begin{figure}
\centerline{\epsfig{file=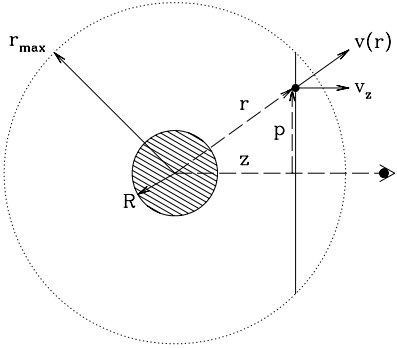,angle=0,width=7cm}}
\caption{\small{A schematic showing the isovelocity zones (surfaces
of constant $v_{z}$) for our velocity model. These are circular
plane surfaces which are oriented normal to the observer's
line-of-sight. We assume that the photosphere is of radius $R$, and
the envelope is of maximum radial extent $r_{\rm max}$.}}
\label{isovelocity}
\end{figure}

\noindent We first need to derive the emission profile for the
unlensed case, expressed in terms of intensity as a function of
frequency and position within the envelope of circumstellar
material. For an envelope in bulk motion such that the flow speed
greatly exceeds the thermal broadening, the locus of points
contributing to the emission at any particular frequency in the line
profile is confined to an ``isovelocity zone'' (c.f.,
Mihalas~\cite{mihalas1978}). These zones are determined by the
Doppler shift formula, namely

\begin{equation}
\nu_{z}=\nu_{0}\left(1-\frac{v_{z}}{c}\right) \, ,
        \label{eq:doppler}
\end{equation}

\noindent where the observer's coordinates are $(x, y, z)$ with the
line-of-sight along $z$, $\nu_{\rm{z}}$ is the Doppler shifted
frequency, and $v_{z} = \bf{-v(r)\cdot\hat{z}}$ is the projection of
the flow velocity onto the line-of-sight (as indicated in Figures
\ref{wind_schematic} and \ref{isovelocity}). Note that both
Cartesian $(x, y, z)$ and spherical $(r, \vartheta, \varphi)$
coordinates can be defined for the star. Employing the Sobolev
theory for line profile calculation in moving media reduces the
radiation transfer in a moving envelope to a calculation in which
one need consider only distinct isovelocity zones.
Equation~(\ref{eq:doppler}) will therefore prove crucial for
relating the variable profile shape to the kinematics of the
envelope.

\noindent For the assumed linear expansion, and for an envelope of
maximum radial extent $r_{\rm{max}}$, the velocity $\bf{v}$ at
radial distance $r$ is given by
\begin{equation}
{\bf{v}} =  v_{\rm{max}} \, (r/r_{\rm{max}}) \, {\bf{\hat{r}}} \, ,
\end{equation}
where $\bf{\hat{r}}$ is a unit vector in the radial direction and
$v_{\rm{max}}$ is the wind speed at $r = r_{\rm{max}}$. The velocity
shift as seen by the observer, situated as in Figure
\ref{wind_schematic}, is then
\begin{equation}
v_{z} = - v_{\rm{max}} (r/r_{\rm{max}})\, \cos \vartheta =
-v_{\rm{max}}\,(z/r_{\rm{max}}) \, , \label{eq:vz}
\end{equation}
where $\vartheta$ is the angle between the radial direction and the
line of sight.

\noindent Thus isovelocity zones for constant $v_{z}$ are seen to be
plane surfaces.  In fact, given the spherical symmetry of our model,
the isovelocity zones are circular plane surfaces of radius
\begin{equation}
p_{\rm{max}} = \sqrt{r_{\rm{max}}^2 - z^2} \, , \label{eq:pmax}
\end{equation}
as depicted in Figure \ref{isovelocity}.

\noindent Consider a projected element of the stellar wind at fixed
impact parameter $p$, and angle $\alpha$ measured clockwise from $x$
-- i.e. $x = p \cos \alpha$ and $y = p \sin \alpha$. The emergent
intensity $I_{\nu}(p)$\footnote{Note that the spherical symmetry of
our model means that the emergent intensity is independent of
$\alpha$.} due to emission from the wind is given by

\begin{equation}
I_{\nu}(p) = \int_z^\infty \, \kappa_{\nu}(r) \,\rho(r)
\,S_{\nu}\,e^{-\tau_\nu}\,dz' \, ,
\end{equation}

\noindent where $\kappa_{\nu}$ is the opacity, $\rho$ the density,
$S_{\nu}$ the source function, and $\tau_{\nu}$ the optical depth at
radial distance $r$, satisfying $r^2 = p^2 + z'^2$.

\noindent The Sobolev approximation is to evaluate the optical depth
{\em only\/} at the point where the Doppler shifted frequency of a
gas parcel exactly equals the observed frequency. This delta
function response in frequency simplifies considerably the
expression for the optical depth, namely

\begin{eqnarray}
\tau_{\nu}(z) & = & \int_{z}^\infty\, \kappa_{\nu}(r) \,\rho(r)\,
\delta(\nu_{z}-\nu_{z'})\,dz' \nonumber \\
              & = & \frac{\kappa_{\nu}(r)\,\rho(r)\,\lambda_0}
{| dv_{z}/dz |_{z} } \, .
\end{eqnarray}

\noindent The denominator on the right hand side is the
line-of-sight velocity gradient.  For a spherical flow, this
gradient is given by the expression

\begin{equation}
\frac{dv_{z}}{dz} = \mu^2\,\frac{dv}{dr} + (1-\mu^2)\,\frac{v}{r} \,
.
\end{equation}

\noindent For the case of homologous expansion, with $v(r) = v_{\rm
max}\,(r/r_{\rm max})$, the line-of-sight velocity gradient
simplifies to $dv_{z}/dz = v_{\rm max}/r_{\rm max}$, which is a
constant. Therefore, although the Sobolev optical depth will
generally depend on angle $\vartheta$ through the factor
$dv_{z}/dz$, homologous expansion is the single exception for which
the optical depth depends on radius only. The optical depth is now

\begin{equation}
\tau_\nu (r) = \frac{\kappa_{\nu}(r)\,\rho(r)\,\lambda_0\, r_{\rm
max}}{v_{\rm max}} \, .
\end{equation}

\noindent Knowing the optical depth as a function of $r$, we can
then obtain an expression for the intensity, $I_{\nu}(p)$, using
equations (4) and (6) and the equation

\begin{equation}
p^2 = r^2 - z^2 \, . \label{eq:prz}
\end{equation}

\noindent In Sobolev theory the intensity then reduces to the form

\begin{equation}
I_{\nu} (p) = S_{\nu}(r)\,\left(1-e^{-\tau_{\nu}(r)}\right) \, .
\end{equation}

\noindent Allowing for line emission from resonance line scattering and
collisional de-excitation, and defining $\epsilon$ as the ratio of the
collisional de-excitation rate to that of spontaneous decay, the
source function can be derived to be

\begin{equation}
S_{\nu} = \frac{\beta_{\rm c}\,I_{*,\nu} +
\epsilon\,B_{\nu}}{\beta+\epsilon} \, .
    \label{eq:Snu}
\end{equation}

\noindent The two parameters $\beta_{\rm c}$ and $\beta$ are
respectively the penetration and escape probabilities, which are given
by

\begin{equation}
\beta_{\rm c} = \frac{1}{4\pi}\,\int_{\Omega_*}\,
    \frac{1-e^{-\tau_\nu}}{\tau_{\nu}}\,d\Omega \, ,
\end{equation}

\noindent and

\begin{equation}
\beta = \frac{1}{4\pi}\,\int_{4\pi}\,
    \frac{1-e^{-\tau_\nu}}{\tau_{\nu}}\,d\Omega \, ,
\end{equation}

\noindent where $\Omega_*$ is the solid angle subtended by the star
at radius $r$.

\noindent The optical depth is generally anisotropic, but as already
noted for homologous expansion the optical depth is isotropic in
this special case.  As a result, the penetration and escape
probabilities are easily calculable, and in particular $\beta_{\rm
c}=W(r)\,\beta$, where the dilution factor
\begin{equation}
W(r) = 0.5\,( 1 - \sqrt{1-R^2/r^2}) \, .
\end{equation}

So for linear expansion, we have the well-known result that the
Sobolev optical depth is a function of radius only.  Since the
source function also depends only on radius (or equivalently on
impact parameter, $p$, at fixed $v_{z}$), so too does the intensity.
To compute the total emergent flux at a given Doppler shift in the
line profile due to emission in the circumstellar envelope, we must
integrate over emergent intensity beams as given by

\begin{equation}
F_{\nu}(v_{z}) = \frac{1}{D^2}\,\int_{v_{z}}\, I_{\nu}(p)\,p\,dp\,
    d\alpha \, , \label{eq:flux}
\end{equation}

\noindent where $D$ is the distance from the Earth. Substituting in
for the intensity, the integration for the flux becomes

\begin{equation}
F_{\nu}(v_{z}) = \frac{2\pi}{D^2}\,\int_{p_{\rm min}}^{p_{\rm max}}
    \,S_{\nu}(r)\,\left(1-e^{-\tau_{\nu}(r)} \right)\,p\,dp \, .
    \label{eq:fnu}
\end{equation}

\noindent where we have, of course, performed the (trivial) integral
over the angle $\alpha$, and the limits $p_{\rm min}$ and $p_{\rm
max}$ are discussed below. For homologous expansion we know that the
Sobolev surfaces are disks oriented transverse to the line-of-sight
(i.e., with $z={\rm constant}$). From equation (\ref{eq:prz}) this
means that $pdp = rdr$, and so the above flux integral could be
equivalently re-formulated as

\begin{equation}
F_{\nu}(v_{z}) = \frac{2\pi}{D^2}\,\int_{r_{\rm min}}^{r_{\rm max}}
        \,S_{\nu}(r)\,\left(1-e^{-\tau_{\nu}(r)} \right)\,r\,dr \, .
\end{equation}

\noindent We now include the contribution of continuum radiation
from a pseudo-photosphere at radius $R$.  It is instructive to
consider separately redshifted and blueshifted wavelengths, which
correspond to $v_{z} > 0$ and $v(z) < 0$ respectively.

\noindent For $v(z) > 0$ (the redshifted side) the total flux
$F_{\rm tot}(v_{z})$ at frequency $\nu$ and line-of-sight velocity
$v_{z}$ may be written as

\begin{equation}
F_{\rm tot}(v_{z}) = F_1(v_{z}) + F_2(v_{z}) \, ,
\end{equation}
where
\begin{equation}
F_1(v_{z}) = \frac{2\pi}{D^2}\,\int_{0}^{R}
        \, I_{\rm phot} \, \, p\,dp \, ,
\end{equation}
and
\begin{equation}
F_2(v_{z}) = \frac{2\pi}{D^2}\,\int_{R}^{p_{\rm max}}
        \,S_{\nu}(r)\,\left(1-e^{-\tau_{\nu}(r)} \right)\,p\,dp \, .
\end{equation}

\noindent Here $I_{\rm phot}$ is the continuum intensity of the
pseudo-photosphere. If $I_{\rm phot}$ is a constant then the
integrals involving $I_{\rm phot}$ become very straightforward.
$F_1$ represents the flux coming directly from the
pseudo-photosphere while $F_2$ accounts for photons scattered by the
surrounding envelope which emerge along the line of sight.  Note
that the lower limit of the $F_2$ integral is $R$ since the region
for $p < R$ is occulted by the pseudo-photosphere of the star.  The
upper limit, $p_{\rm max}$ satisfies the relation
\begin{equation}
p_{\rm max}^2 = r_{\rm max}^2 - z^2 \, . \label{eq:przmax}
\end{equation}

\noindent For $v(z) < 0$ (the blueshifted side) the total flux
consists of three terms, i.e.
\begin{equation}
F_{\rm tot}(v_{z}) = F_3(v_{z}) + F_4(v_{z}) + F_5(v_{z}) \, ,
\end{equation}
where
\begin{equation}
F_3(v_{z}) = \frac{2\pi}{D^2}\,\int_{0}^{p_{\rm lim}}
        \, I_{\rm phot} \, \,p\,dp \, ,
\end{equation}
\begin{equation}
F_4(v_{z}) = \frac{2\pi}{D^2}\,\int_{p_{\rm lim}}^{p_{\rm max}}
        \,S_{\nu}(r)\,\left(1-e^{-\tau_{\nu}(r)} \right)\,p\,dp \, ,
\end{equation}
and
\begin{equation}
F_5(v_{z}) = \frac{2\pi}{D^2}\,\int_{p_{\rm lim}}^{R}
        \, I_{\rm phot} \, e^{-\tau_{\nu}(r)} \, p\,dp \, .
\end{equation}

\noindent Here $p_{\rm lim}$ is defined as
\begin{equation}
p_{\rm lim} = \left\{ \begin{array}{ll}
 \sqrt{R^2 - z^2} & \quad {\rm for} \, \, 0 \leq z \leq R \\
 0 & \quad {\rm for} \, \, z > R
\end{array} \right. \quad .
\end{equation}

\noindent Note that if $z > R$ then the $F_3$ integral is
identically zero; in this case the flux contribution from the
pseudo-photosphere is represented fully by the $F_5$ term, which
also takes account of attenuation by intervening wind material along
the line-of-sight.

\noindent Finally we must stress that, in accordance with the
standard notation adopted in the microlensing literature, all
distance scales are in fact taken as angular distances that are
normalised to the angular Einstein radius of the lens. This implies
that the ``fluxes'' defined in the above equations have rather
unusual units. however, the results of our line profile calculations
in the following sections will be displayed as ratios normalised by
the continuum, so that the (non-standard) units of flux will cancel.

\section{Unlensed profiles from homologous expansion}
\label{sec:unlensed}

Before including the effects of gravitational lensing we first
consider the unlensed profiles derived by applying the model of the
previous section to the illustrative cases of a pure resonance line
and a pure recombination line.  We have computed line profiles for a
range of optical depths, assuming a spherically symmetric density
distribution for the ejecta given by

\begin{equation}
\rho (x) = \rho_0\, x^{-3} \, , \label{eq:rho}
\end{equation}

\noindent where $x=r/R$.  The factor $x^{-3}$ accounts for mass
continuity within the shell as material expands.  The scale
parameter $\rho_0$ denotes the density at the radius of the
pseudo-photosphere and specifies the total amount of gas ejected in
the explosion, via the equation

\begin{equation}
M_{\rm ejecta} = \int \, 4\pi r^2\,\rho(r) \, dr \, .
\end{equation}

\subsection{Pure resonance lines}

For the case of a pure resonance line the parameter $\epsilon$ is
set to zero in eq.~(\ref{eq:Snu}), and the source function becomes

\begin{equation}
S_{\nu} = W(r)\,I_{*,\nu} \, .
\end{equation}

\noindent The opacity for resonance line scattering does not depend
explicitly on density, but an implicit dependence can arise through,
for example,  the ionization fraction of whatever atomic species is
being considered.  For simplicity, we suppress any radial dependence
of the opacity in our model, and parametrize the Sobolev optical
depth as

\begin{equation}
\tau_{\nu} = \tau_0\,(\rho/\rho_0) = \tau_0 \, x^{-3}  \, .
\label{eq:tau}
\end{equation}

\noindent An especially convenient parameter used to characterize
different line calculations is the line integrated optical depth
defined as follows

\begin{equation}
{\rm T}  =  \int^{x_{\rm{max}}}_1\,\tau_{\nu}\,dx \, = \,
\frac{1}{2} \tau_0 \left ( 1 - \frac{1}{x_{\rm max}^2} \right ) \, .
\label{eq:tres}
\end{equation}

\noindent Thus, for fixed $x_{\rm{max}}$, it is trivial to compute
the value of $\tau_0$ -- and hence the model profile --
corresponding to a given value of $\rm T$.

\noindent Figure~\ref{resun} shows some examples of model profiles
computed for $\rm T = 0.3, 1.0, 3.0$ and $10.0$, with stronger lines
corresponding to larger values of $\rm T$. In all cases a fixed
value of $x_{\rm max} = 5$ was adopted, so that from equation
(\ref{eq:tres}) $\rm T = 0.48 \tau_0$. Note that our chosen value of
$x_{\rm max}$ is responsible for the change of slope of each line
profile at $v_z = -0.2 v_{\rm max}$. This is due to the occultation
of wind material which lies directly behind the pseudo-photosphere
of the star (at $x = 0.2 x_{\rm max}$ in this case).

\noindent Increasing the value of $x_{\rm max}$ would result in
qualitatively similar line profiles but with a narrower emission
peak and absorption trough. We can understand this behaviour as
follows.  From equations (\ref{eq:vz}) and (\ref{eq:prz}) we can
write

\begin{equation}
x = \sqrt{ \frac{p^2}{R^2} + w^2 x_{\rm max}^2} \, \, ,
\end{equation}
where $w = v_z / v_{\rm max}$. From equation (\ref{eq:tau}) it then
follows that for $w \simeq 0$ the optical depth -- and hence the
line flux -- is insensitive to $x_{\rm max}$ (The \emph{range} of
integration in equation (\ref{eq:fnu}) increases as $x_{\rm max}$
increases, but this has little effect on the line flux because the
optical depth is proportional to $x^{-3}$).

\noindent For $w \simeq \pm 1$, on the other hand, at fixed $p$ the
optical depth decreases sharply as $x_{\rm max}$ increases --
resulting in reduced line emission and absorption and thus a
narrowing of these features in the line profile.

\subsection{Pure recombination lines}

We have also computed model profiles for the case of a pure
recombination line.  As such, scattering is ignored, and the source
function is assumed to arise from an LTE process, so that

\begin{equation}
S_{\nu} = B_{\nu}(T) \, ,
\end{equation}

\noindent for temperature, $T$. For simplicity, the envelope shall
be taken as isothermal. The opacity for a recombination line is
$\kappa \propto \rho$, and so the scaling of the optical depth will
for this case be

\begin{equation}
\tau_{\nu} = \tau_0\,(\rho/\rho_0)^2.
\end{equation}

\noindent As a $\rho^2$ process, the optical depth will go as
$x^{-6}$ -- a strong function of radius.

\noindent It is again useful to introduce the integrated optical
depth parameter given by

\begin{equation}
{\rm T} = \int^{x_{\rm max}}_1\, \tau_0\,(\rho/\rho_0)^2\,dx \, ,
\label{eq:trec}
\end{equation}
so that again, for fixed $x_{\rm max}$, it is trivial to compute the
value of $\tau_0$ corresponding to a line of given integrated
optical depth.

\begin{figure}
\centerline{\epsfig{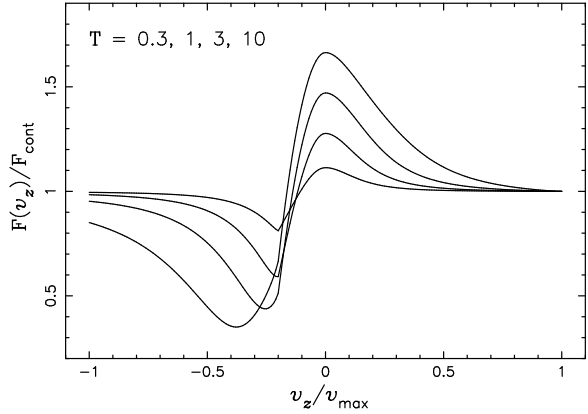}}
\caption{\small{Model line profiles from pure resonance line
scattering, calculated using the method described in the text, for
four values of the integrated optical depth, $\rm T$. The stronger
lines correspond to larger values of $\rm T$.}} \label{resun}
\end{figure}

\begin{figure}
\centerline{\epsfig{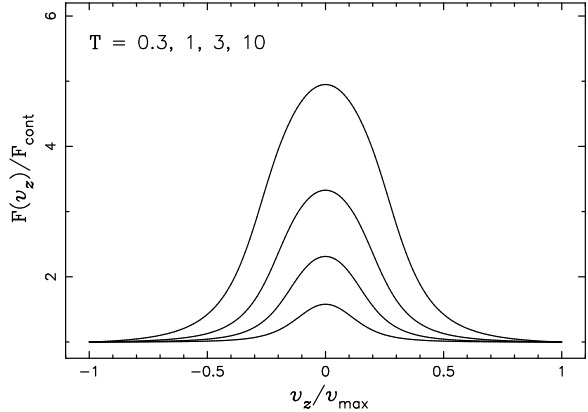}}
\caption{\small{Model line profiles from pure recombination line
scattering, calculated using the method described in the text, for
four values of the integrated optical depth, $\rm T$. The stronger
lines correspond to larger values of $\rm T$.}} \label{recun}
\end{figure}

\noindent Figure~\ref{recun} shows profile shapes again computed for
$x_{\rm max}=5$ and for the same range of values of $\rm T$ as in
Figure~\ref{resun}. From equation (\ref{eq:trec}) it follows that
$\rm T \approx 0.2\tau_0$. Note that in contrast to
Figure~\ref{resun}, the recombination lines do not exhibit any
blueshifted absorption. That is because we assume that the continuum
emission is given by a blackbody with the same temperature as the
envelope. Consequently for sightlines that intercept the continuum
forming surface, one has $I_\nu = B_\nu e^{-\tau} +
B_\nu(1-e^{-\tau}) = B_{\nu}$.  Hence the combined emission from
line plus continuum will always equal or exceed the continuum level
when integrated over all sightlines.

\noindent As was the case for pure resonance lines, increasing
$x_{\rm max}$ would produce narrower profiles but would have
negligible effect on the value of the peak emission at $w = 0$ for
fixed $\rm T$.

\section{Lensed profiles from homologous expansion}

We now consider the effect of microlensing on emission line profiles
for a stellar wind in homologous expansion.  Our expression in
equation (\ref{eq:flux}) for the emergent flux is now replaced by an
integral over emergent intensity beams weighted by the
(dimensionless) lensing magnification $A(d)$, where $d$ is the
projected separation between the lens and an area element of the
wind in the source plane, i.e.

\begin{equation} F(v_z) = \frac{1}{D^2} \,
\int_{0}^{2\pi} \, \int_{p_{\rm min}}^{p_{\rm max}} \, I_\nu (p)A(d)
\, p \, dp\, d\alpha \, \, .
\end{equation}

\noindent Note that, while we continue to assume that $I_\nu$
depends only on impact parameter, the lensing magnification is a
function of both $p$ and $\alpha$, so that we must perform a double
integration to compute the emergent flux at each radial velocity.
Moreover, since the magnification pattern of the lens evolves with
time, as the relative projected positions of the lens and wind
change, the computed line profile will also be time-dependent.

\noindent The form of the magnification function is determined by
the nature of the lens.  In this paper we consider two illustrative
cases:  those of a point mass lens and a fold caustic.

\subsection{Point mass lensing event}
In the typical microlensing situation of a point mass lens the
(time-dependent) magnification factor takes the well-known form (see
e.g. Pacy\'nski 1986)
\begin{equation}
A(u)=\frac{u^2+2}{u\sqrt{u^2+4}}
\end{equation}
where $u$ is the {\em impact parameter\/}: the projected separation
between the lens and source element, given by
\begin{equation}
u(t) = \sqrt{u_0^2 + \frac{(t-t_0)^2}{t_E^2}} \label{eq:uoft}
\end{equation}
where $t_0$ is the time of closest approach between lens and source
element, which occurs at $u = u_0$, the minimum impact parameter,
and $t_E$ is the characteristic timescale of the event. Both $u$ and
$u_0$ are expressed in units of the angular Einstein radius of the
lens, which is defined as

\begin{equation}
\theta_E = \sqrt{\frac{4GM}{c^2} \frac{D_S - D_L}{D_S D_L}}
\label{eq:aer} \end{equation} where $M$, $D_S$, $D_L$ are the lens
mass, source distance and lens distance respectively.  The timescale
$t_E$ may be written as

\begin{equation}
t_E = \frac{\theta_E}{\mu_{\rm rel}} \label{timescale}
\end{equation}
where $\mu_{\rm rel}$ is the relative proper motion of the lens.

\noindent In the limit of large separations the magnification is, of
course, equal to unity while for small $u$ the magnification goes
approximately as $1/u$. Thus if a point lens transits an extended
source, while formally the magnification factor is infinite for the
source element directly behind the lens, when one integrates over
the finite area of the source an finite value for the magnification
is obtained.

\begin{figure}
\centerline{\epsfig{file=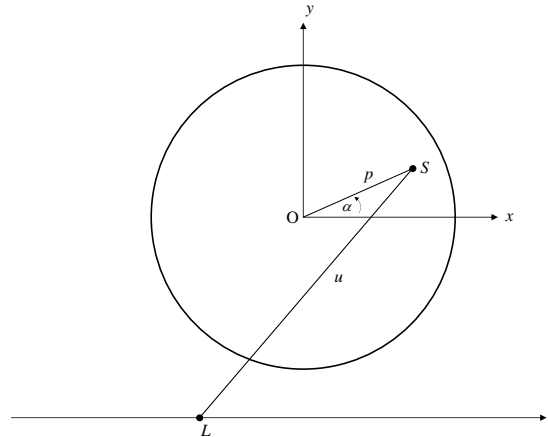,angle=0,width=7.7cm}}
\caption{\small{A schematic view of the source plane for a point
lens event, showing in projection (i.e. with the observer in the $z$
direction) the outer edge of the circumstellar envelope (of radius
$r_{\rm max}$) and the trajectory of the point lens, $L$. An element
of the source, at $S$, with position defined by polar coordinates
$p$ and $\alpha$, is shown; it lies at a projected separation $u$
from the lens.}} \label{point}
\end{figure}

\noindent Figure \ref{point} illustrates schematically the geometry
of a point lens event. The outer edge of the circumstellar envelope
(of radius $r_{\rm max}$) is shown in projection, together with the
trajectory of the lens.  When the lens is at position $L$, an area
element of the wind at position $S$ (defined by polar coordinates
$p$ and $\alpha$ -- see eq. 36) lies a projected distance $u$ from
$L$.

\begin{figure*}
\centerline{\epsfig{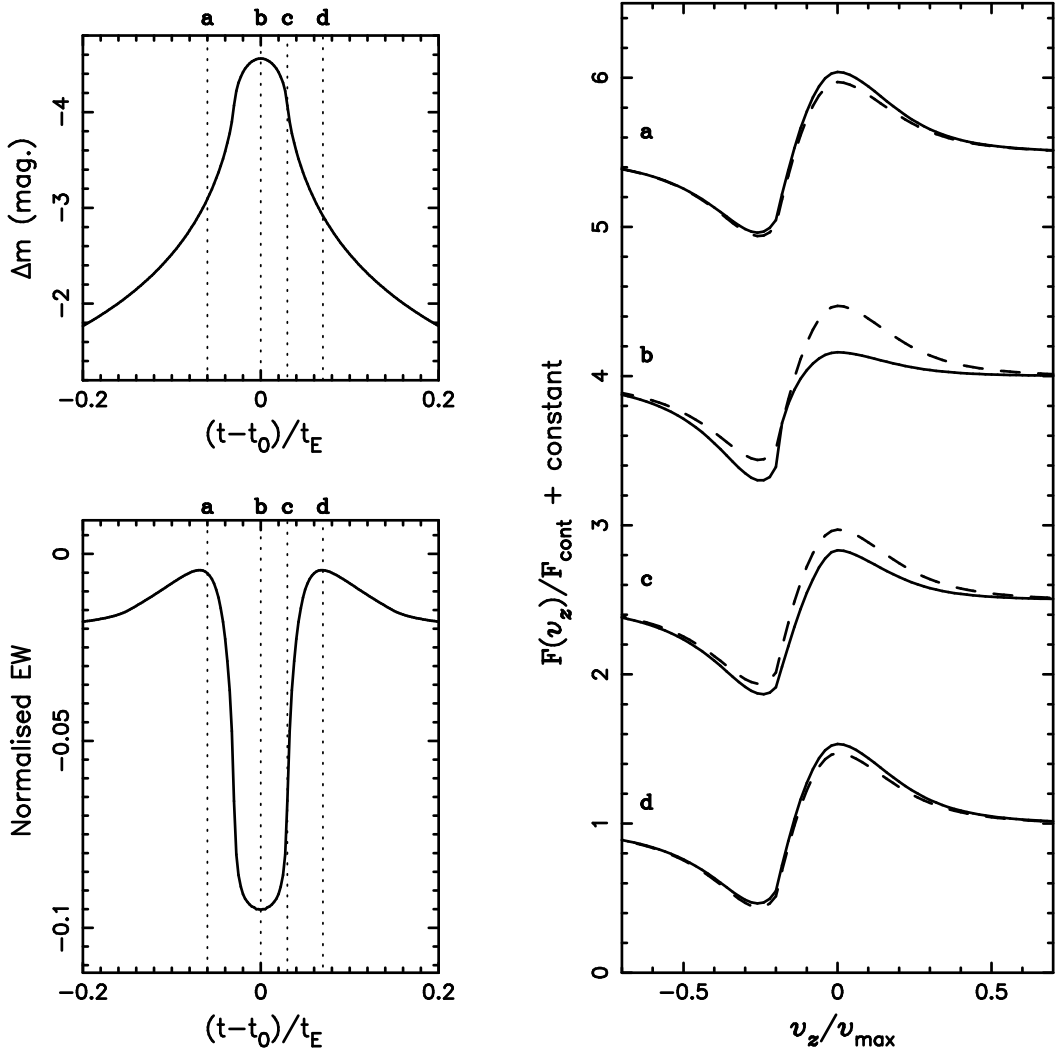}}
\caption{\small{The lightcurve, equivalent widths and line profiles
produced by a point mass lens transiting a star with photospheric
radius $R=0.03\,\theta_E$ surrounded by a circumstellar envelope
exhibiting pure resonance line scattering. The upper left panel
shows the change in apparent magnitude of the lensed continuum flux,
compared with the unlensed flux, as a function of time.  Below it
the time evolutioin of the normalised equivalent width of the
resonance line is shown. The four line profiles in the right hand
panel correspond to the epochs represented by dotted vertical lines
in the right hand panel. Thus, for example, profile `b' corresponds
to the epoch of maximum magnification (when the lens is at its
minimum impact parameter) and is also when the largest change in
equivalent width occurs. The solid lines in the right hand panel
show the lensed line profiles while the dashed lines show the
unlensed profiles, and are included for comparison. The offset
between the profiles is provided for clarity. All profiles were
calculated for an integrated optical depth of $T=3.0$ and assuming
$x_{\rm max} = 5$.}} \label{plres}
\end{figure*}

\begin{figure*}
\centerline{\epsfig{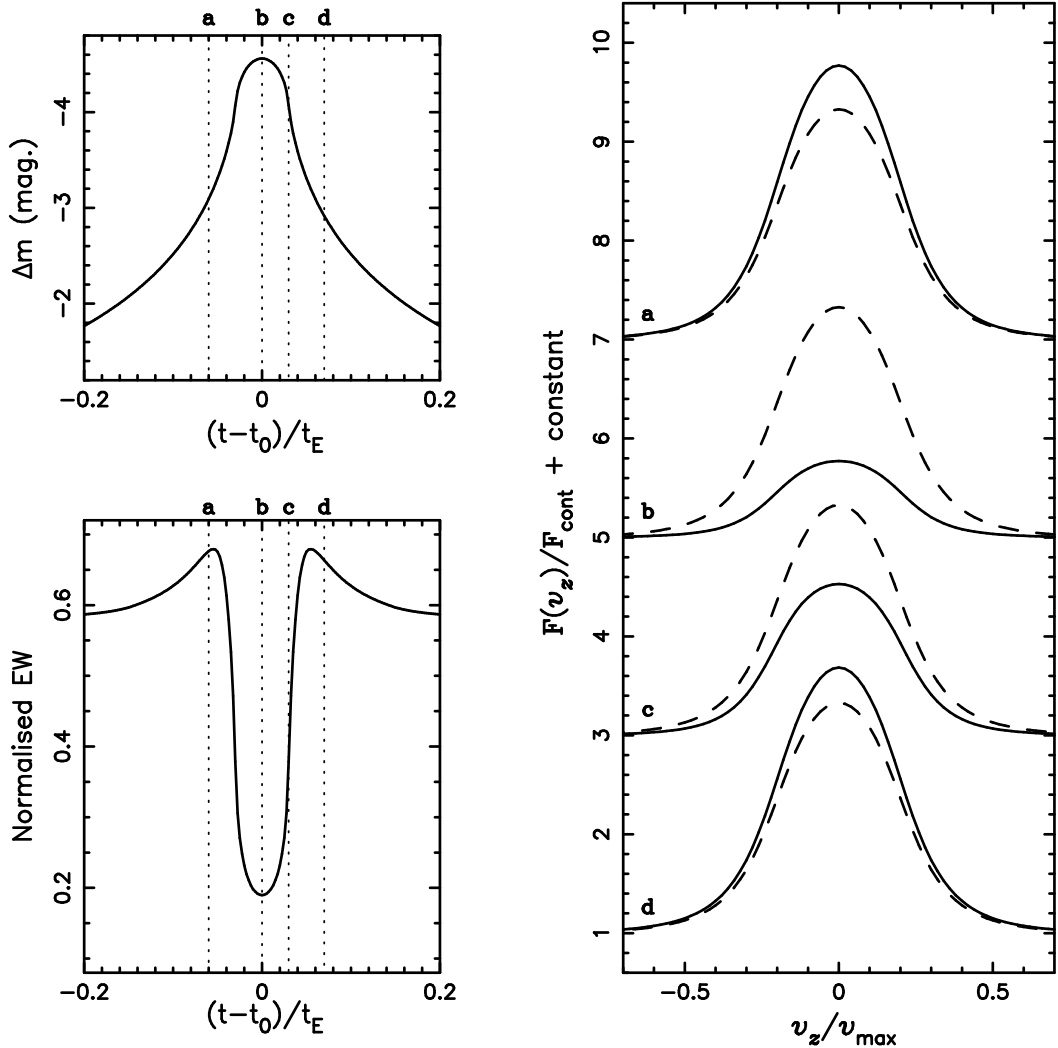}}
\caption{\small{The lightcurve, equivalent widths and line profiles
produced by a point mass lens transiting a star with photospheric
radius $R=0.03\,\theta_E$ surrounded by a circumstellar envelope
exhibiting pure recombination line scattering. The three panels
present results in the same manner as in Figure \ref{plres}. All
profiles were again calculated for an integrated optical depth of
$T=3.0$ and assuming $x_{\rm max} = 5$.}} \label{plrec}
\end{figure*}

\subsubsection{Pure resonance lines}

We computed continuum lightcurves, line equivalent widths and line
profiles for the case of a pure resonance line lensed by a point
mass lens.  Illustrative results are shown in Figure \ref{plres} for
a transit with minimum impact parameter equal to zero -- i.e.
assuming the lens crosses exactly over the centre of the
circumstellar envelope\footnote{In view of the symmetry of our model
the orientation of the lens trajectory during transit is irrelevant}
-- and adopting a photospheric radius $R = 0.03 \, \theta_E$.  This
value is representative of source radii determined for observed
microlensing events exhibiting finite source effects. Indeed Macho
event M95-30, which was the first well-documented example of a
finite source event, was found to have a somewhat larger radius of
$R \approx 0.075 \, \theta_E$ (Alcock et al. 1997). In the results
shown in Figure \ref{plres} we truncated the radius of the
circumstellar envelope at $x_{\rm max} = 5$ and the line profiles
were taken to have integrated optical depth $\rm T = 3$.

\noindent The upper left hand panel of Figure \ref{plres} shows the
continuum lightcurve of the lensing event -- i.e. the change in
apparent magnitude, computed from the ratio of lensed to unlensed
continuum flux, as a function of time.  The time axis is measured in
units of $t_E$ -- the lensing timescale introduced in equation (40).
As expected, the maximum magnitude change occurs at minimum impact
parameter, when the lens transits the centre of the photosphere, and
the continuum flux is boosted by a factor of more than 60.  Only the
central portion of the light curve is shown; although lensing still
produces a continuum flux magnification of $\sim 5$ when
$(t-t_0)/t_E = \pm 0.2$, the effect of lensing on the line profiles
is essentially negligible for larger impact parameters.

\noindent The lower left hand panel of Figure \ref{plres} shows the
time evolution in the normalised equivalent width of the resonance
line. The equivalent width was calculated according to the formula

\begin{equation}
{\rm EW} = \left < \frac{F(v_z)}{F_{\rm cont}}  \right > \, - \, 1
\, \, , \label{eq:ew}
\end{equation}
where the angled brackets denote the mean, or expectation, value
obtained by integrating over $v_z$.

\noindent The vertical dashed lines labelled a, b, c and d in the
upper and lower left hand panels correspond to the four epochs in
the right hand panel at which the lensed (solid curve) and unlensed
(dashed curve) line profiles are compared. For clarity a constant
vertical offset is introduced between each of the four epochs; note
that this does not indicate any change in magnification.

\noindent  We can see from the lower left panel that the unlensed
value (i.e. for large impact parameter) of the the normalised
equivalent width is slightly negative because of the asymmetry in
the unlensed line profile:  the area of the blueshifted absorption
trough is slightly larger than that of the redshifted emission peak.
As the event proceeds and the lens crosses the outer edge of the
envelope, at $(t-t_0)/t_E = -0.15$, the normalised equivalent width
at first shows a clear increase, as the lens differentially
magnifies a portion of the isovelocity zones in emission. This stage
of the lensing event corresponds to profile (a) in the right hand
panel. As the projected lens position approaches the photosphere,
however, the absorption component of the line profile quickly begins
to dominate.  This is because the lens is differentially magnifying
the attenuated continuum emission coming from the photosphere.  The
equivalent width reaches its lowest value when the lens has zero
impact parameter -- i.e. when it lies directly in front of the
centre of the photosphere. The line profile at this epoch is shown
in the right hand panel as profile (b).  The lensing event
thereafter is a mirror image of what is observed on the incoming
trajectory: the equivalent width rises sharply as the lens exits the
photosphere, reaches a peak value close to the edge of the
circumstellar envelope before tending to its unlensed value at large
impact parameter. The two further line profiles at epochs (c) and
(d) illustrate this behaviour.

\subsubsection{Pure recombination lines}

We also computed continuum lightcurves, equivalent widths and line
profiles for the case of a pure recombination line lensed by a point
mass lens. Illustrative results are shown in Figure \ref{plrec} for
exactly the same wind and lens parameters as in Figure \ref{plres}.

\noindent The upper left panel of Figure \ref{plrec} is, of course,
identical to that of Figure \ref{plres}, since it shows the
magnification of the continuum flux.  The time evolution of the
equivalent width also shows qualitatively similar behaviour to that
of the resonance line case, although the magnitude of the maximum
change in equivalent width is approximately five times larger; this
is essentially due to the stronger dependence of the optical depth
on radius for a recombination line.

\noindent Again we see that the equivalent width first increases as
the lens transits the circumstellar envelope, due to the
differential magnification of the emission flux from the region
around the star. When the lens transits the photosphere, on the
other hand, the differential magnification of the attenuated
continuum flux from the photosphere again causes a sharp decrease in
equivalent width, which reaches its minimum value at zero impact
parameter. The symmetry of the isovelocity zones and the lens
trajectory then results in symmetric behaviour during the second
half of the event.

\subsection{Fold caustic lensing event}

For a binary lensing event the magnification pattern in the source
plane is, in general, a rather complicated function of the lens
masses and separations. As we remarked in Section 4.1, for a single
point lens the magnification $A(u)$ goes as $1/u$ as $u \rightarrow
0$, so that the magnification is formally infinite at only one point
-- coincident with the position of the point lens itself, for which
$u = 0$.  For a binary lens, on the other hand, the magnification is
formally infinite along one or more closed curves, or {\em caustic
structures\/}, in the plane of the sky. Any source crossing into or
out of a caustic will experience a high degree of magnification, and
-- no matter how small -- must be modelled as an extended source;
i.e. the total magnification must be computed by dividing the source
into (formally infinitesimal) area elements, calculating the
magnification experienced by each area element, and summing (or,
formally, integrating) the results. Moreover, as a source (or an
area element thereof) enters a caustic structure, two extra images
of it are produced by the binary lens. These extra images cause a
sudden and sharp increase in the total magnification.

\noindent If the angular size of the source is very small compared
with that of the closed caustic curve, then generally we may
approximate by a straight line the portion of the caustic curve in
the vicinity of the source. This situation is referred to as the
{\em fold caustic approximation\/}, and the magnification function
in this case can be adequately described as (Schneider et al. 1992)

\begin{equation}
A(d) = \cases{ A_{0}+\frac{b_{0}}{\sqrt{d}} & ($d<0$), \cr  A_{0} &
($d>0$). } \label{eq:caustamp}
\end{equation}
Here $A_0$ is the total magnification of the 3 images which form
when the source is {\em outside\/} the caustic structure; to a very
good approximation this term is constant during the caustic
crossing.  The distance $d$ is the projected source--caustic
separation (i.e. the perpendicular distance between the source and
the line representing the fold caustic) in units of the angular
Einstein radius of the binary lens (which is given by eq.
\ref{eq:aer} as for a point lens, but with the mass $M$ equal to the
combined mass of both binary components). The constant $b_0$ depends
on the parameters of the lens system, but is of order unity for
typical caustics, and in our calculations for simplicity we set it
exactly equal to unity.

\begin{figure}
\centerline{\epsfig{file=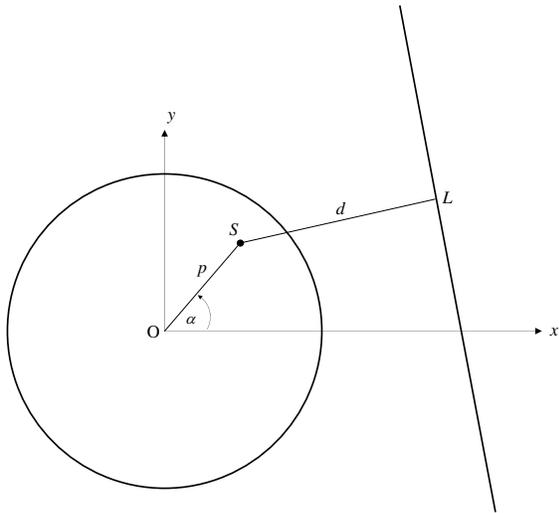,angle=0,width=7.7cm}}
\caption{\small{A schematic view of the source plane for a fold
caustic crossing event, showing in projection (i.e. with the
observer in the $z$ direction) the outer edge of the circumstellar
envelope (of radius $r_{\rm max}$).  The bold line represents a
small portion of a closed caustic structure, approximated by a
straight line.  An element of the source, at $S$, with position
defined by polar coordinates $p$ and $\alpha$, is shown; it lies at
a projected perpendicular distance $d$ from the fold caustic.  Note
that the fold caustic will not, in general, be aligned with the
direction of motion of the source (assumed here to be along the
$x$-axis) so that the caustic may sweep across the source at an
oblique angle.}}\label{fold}
\end{figure}

\noindent Note that the sign of $d$ is important here: the excess
magnification from the extra two images occurs only when the source
(or an element thereof) lies {\em inside\/} the caustic structure.
In equation (\ref{eq:caustamp}), therefore, the magnification is
defined according to the sign convention that $d$ increases from
left to right, and the caustic interior lies to the left of the fold
caustic (i.e. for $d<0$). This situation is illustrated
schematically in Figure \ref{fold}, which shows in projection the
outer edge of the circumstellar envelope (of radius $r_{\rm max}$)
and a small portion of a closed caustic structure, approximated by a
straight line.  An element of the source, at $S$, with position
defined by polar coordinates $p$ and $\alpha$, is also shown; it
lies at a projected perpendicular distance $d$ from the fold
caustic.  Note that the fold caustic will not, in general, be
aligned with the direction of motion of the source (assumed in Fig.
\ref{fold} to be along the $x$-axis) so that the caustic may sweep
across the source at an oblique angle. This oblique case is shown in
Figure \ref{fold}; in the examples presented in the next two
sections, however, we consider only the case where the fold caustic
is perpendicular to the direction of motion of the source.

\noindent Note further that the fold caustic approximation will
generally be good, since the size of the caustic structure will
typically be much larger than the size of the source and hence
curvature of the caustic will not be important. The approximation
will break down if the caustic crossing occurs close to a `cusp' in
the caustic structure, but we do not consider that case here.  We
will extend our treatment to more general caustic models (and to
other expansion laws) in future work.

\subsubsection{Pure resonance lines}

\begin{figure*}
\centerline{\epsfig{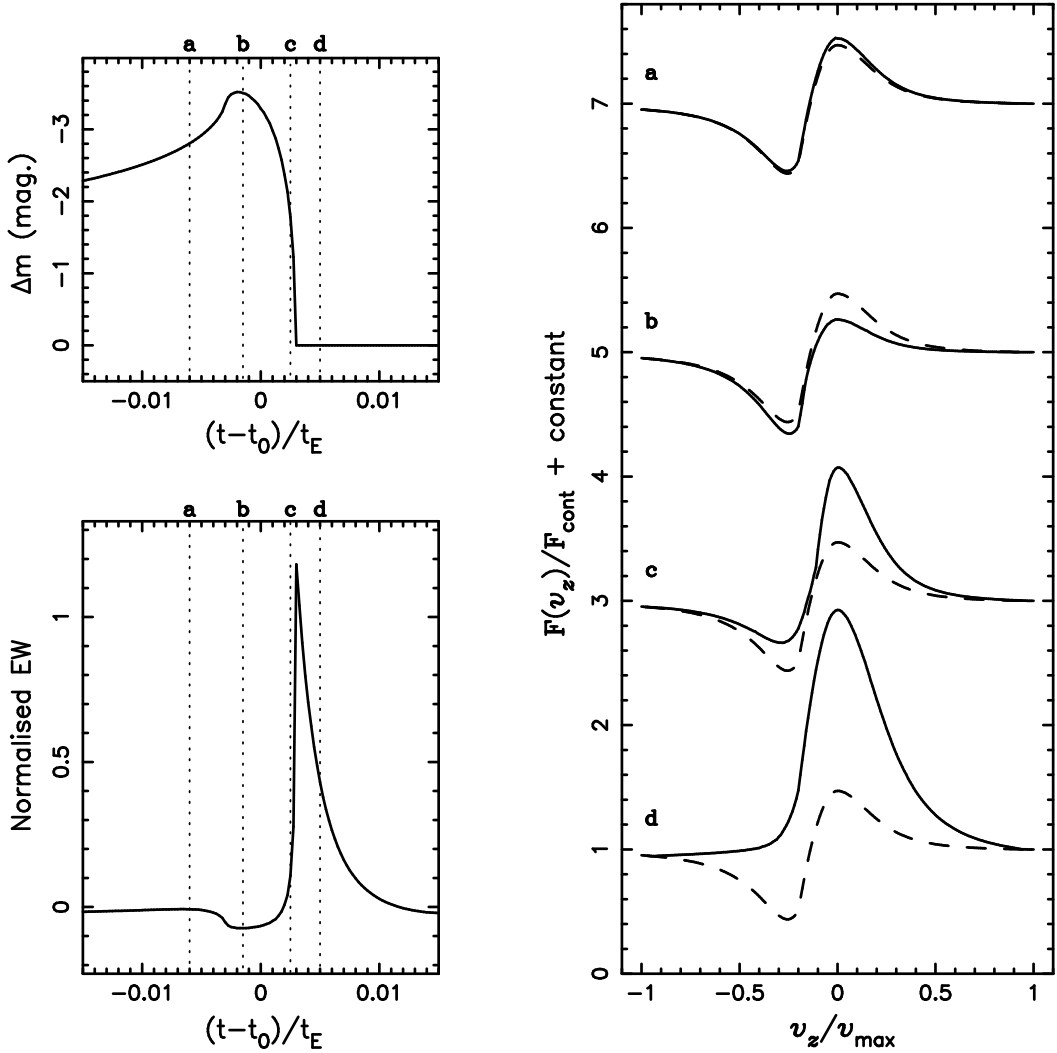}}
\caption{\small{The lightcurve, equivalent widths and line profiles
produced by a fold caustic crossing of a star with photospheric
radius $R=0.003\,\theta_E$ surrounded by a circumstellar envelope
exhibiting pure resonance line scattering. The three panels present
results in the same manner as in Figures \ref{plres} and
\ref{plrec}. All profiles were again calculated for an integrated
optical depth of $T=3.0$ and assuming $x_{\rm max} = 5$.}}
\label{lcres}
\end{figure*}

\begin{figure*}
\centerline{\epsfig{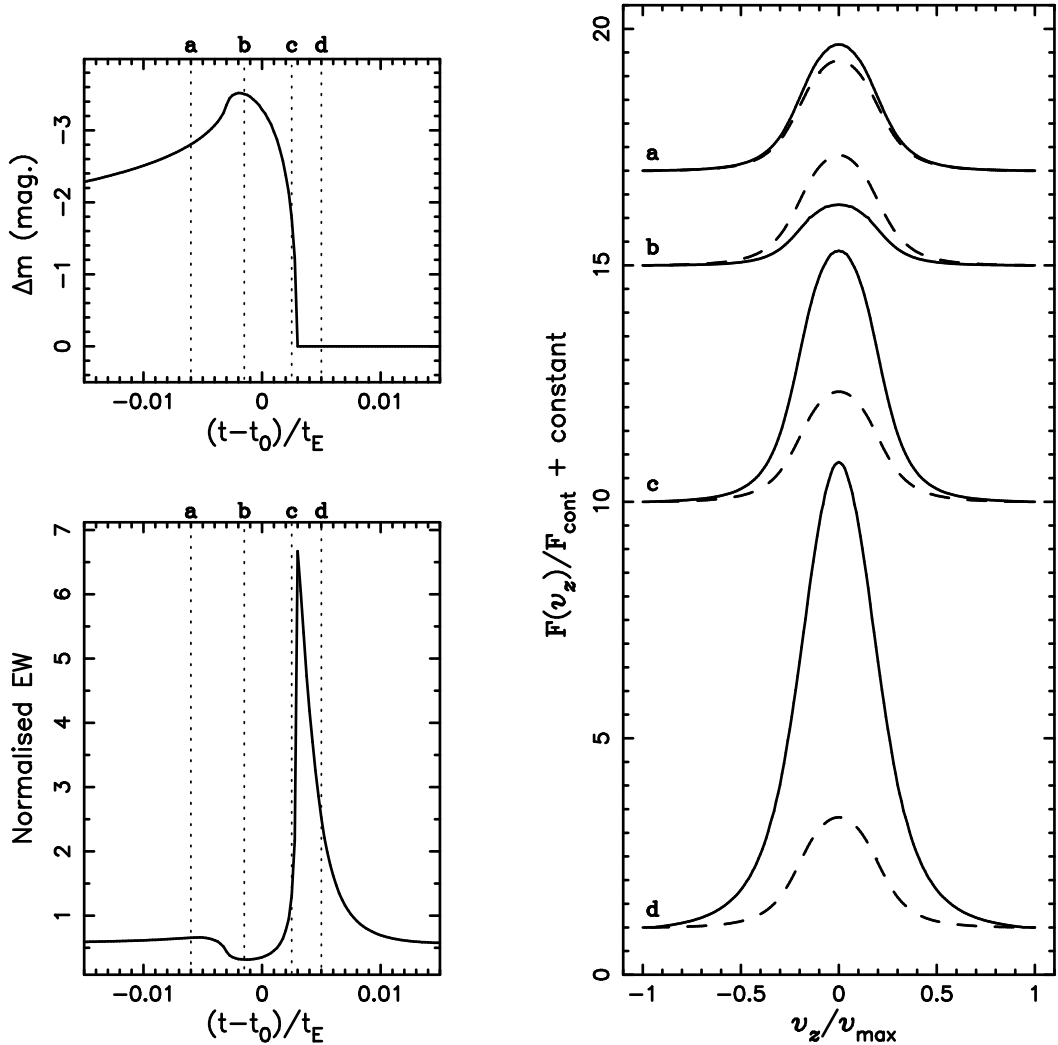}}
\caption{\small{Same as for Figure \ref{lcres}, but now shown for a
pure recombination line.}} \label{lcrec}
\end{figure*}

We next computed continuum lightcurves, line equivalent widths and
line profiles for the case of a pure resonance line lensed by a fold
caustic as the source exits the caustic structure. Illustrative
results are shown in Figure \ref{lcres} for a star with
pseudo-photospheric radius $R = 0.003 \, \theta_E$ and assuming
$x_{\rm max} = 5$.  The model line profile shown has $\rm T = 3$.

\noindent The upper left hand panel of Figure \ref{lcres} shows the
change in apparent magnitude, computed from the ratio of lensed to
unlensed continuum flux, as a function of time (again measured in
units of $t_E$).  Note that we consider here only the effect of the
excess magnification due to the extra images produced by the caustic
structure; in other words we are assuming that the term $A_0$ in
equation (\ref{eq:caustamp}) is a constant during the caustic
crossing, and can therefore be subtracted from the total
magnification, leaving only the differential effect of the two extra
images. Hence the magnitude change due to the extra images alone
drops to zero after the photosphere has fully exited the fold
caustic.

\noindent The lower left hand panel of Figure \ref{lcres} shows the
time evolution of the line equivalent width, calculated as before.
We see that the equivalent width first rises gently as the fold
caustic begins to cross the circumstellar envelope and
differentially magnifies the line emission from the wind relative to
the continuum.  The increase in equivalent width is at first modest,
however, because the absorption components of the line profile also
receive some differential magnification at this stage.  Indeed once
the fold caustic begins to cross the star, the dominant effect of
the differential magnification is to enhance the attenuated
continuum flux from the pseudo-photosphere -- thus resulting in a
sharp dip in the equivalent width which reaches its minimum value at
epoch (b). Thereafter the equivalent width begins to rise
dramatically. This is because an increasing fraction of the
pseudo-photosphere now lies {\em outside\/} the caustic structure,
so that the relative effect of the lens on the circumstellar
material which still lies inside is significantly enhanced. This
effect reaches its peak shortly after epoch (c), by which time the
photosphere lies {\em entirely\/} outside the caustic but a portion
of the circumstellar envelope still lies inside, and thus continues
to be differentially magnified due to the existence of the two extra
images. As the fold caustic then sweeps across the rest of the
envelope, the equivalent width drops sharply again, although at
epoch (d) the equivalent width is still somewhat larger than at
epoch (c) -- as is evident in the profiles shown in the right hand
panel.

\subsubsection{Pure recombination lines}

Finally we computed continuum lightcurves, line equivalent widths
and line profiles for the case of a pure recombination line lensed
by a fold caustic as the source exits the caustic structure.
Illustrative results are shown in Figure \ref{lcrec} for eaxctly the
same lens and source parameters as in Figure \ref{lcres}.

\noindent The upper left panel of Figure \ref{lcrec} is again
identical to that of Figure \ref{lcres}. From the lower left panel
we see that the time evolution of the line equivalent width shows
qualitatively similar behaviour to that of Figure \ref{lcres},
although again the magnitude of the change in equivalent width is
significantly larger for a recombination line than for a resonance
line -- as we also found for a point lens event.  The stronger
signature of lensing is also clearly seen in the right hand panel of
Figure \ref{lcrec}, where the lensed line profiles at epochs (a) to
(d) show significantly greater enhancement of the emission near line
centre than the enhanced emission at the corresponding epochs of
Figure \ref{lcres}, or in Figure \ref{plrec} for a recombination
line lensed by a point mass lens.

\noindent Thus we can see collectively from these illustrative
figures that the time evolution of the equivalent width as the
lensing event proceeds can be an effective diagnostic of the radial
extent of the circumstellar envelope.  The equivalent width
variations will also be sensitive to the integrated optical depth of
the line; as we saw in Figures \ref{resun} and \ref{recun}
increasing $\rm T$ resulted in a stronger line. Therefore, by
studying the time evolution of equivalent width for a number of
different lines, microlensing can in principle be a powerful probe
of the run of optical depth as a function of radius in the
circumstellar envelope.

\section{Discussion}

In this paper we have modelled the effects of microlensing for
variations of line profiles formed in circumstellar envelopes.
Although the microlensing is quantitatively accurate, the underlying
wind model includes a number of simplifying assumptions: the flow is
assumed to be spherically symmetric; resonance and recombination
lines are both treated, but without detailed consideration of
ionization or temperature gradients in the flow; the wind velocity
law is treated as a linear function of radius; the envelope is
truncated at some maximum radius; Sobolev theory is employed for the
line formation.  The intention of our modeling has been to assess
relative effects from the line variations for an illustrative model,
one that has generic application to stellar winds, but that would
nonetheless require modification when applied to any particular
case.

\noindent At this juncture it is useful to comment on the diverse
kinds of winds observed across the Hertzsprung-Russell Diagram
(hereafter, ``HRD'').  Referring to the introductory stellar winds
book by Lamers \& Cassinelli (1999), most stellar winds fall into
two primary classes, these being radiatively driven or pressure
driven.  Of the former, line-driven winds are the primary
consideration for hot OB stars (e.g., see review by Kudritzki \&
Puls 2000)  as basically described by the CAK theory of Castor,
Abbott, \& Klein (1975).  Line-driving is also relevant for the
central stars of planetary nebulae that show Wolf-Rayet-like spectra
(Tylenda, Acker, \& Stenholm 1993).  The Main Sequence hot stars
exhibit absorption line spectra from their photospheres; however,
strong and weak P~Cygni lines are abundant at UV wavelengths
(Howarth \& Prinja 1989; Snow et al. 1994). In the evolved hot
stars, recombination lines appear at IR wavelengths (e.g., Lenorzer
et al. 2002). Also, a substantial fraction (around a quarter,
although the value is uncertain) of B stars are classified as
emission-line or ``Be'' stars (Porter \& Rivinius 2003). These have
circumstellar disks that produce copious recombination lines in the
IR, but more interestingly show H$\alpha$ in emission, which would
be relevant for ground-based follow-up monitoring programs.
Unfortunately, hot stars are relatively rare and so are extremely
unlikely to be background sources for microlensing events.

\noindent Working toward cooler temperatures in the HRD, the winds
from Main Sequence stars A--M are not well-studied, in part because
their mass loss is so much weaker than early-type stars.  The yellow
hypergiants have more substantial winds, but these stars are
extremely rare (de Jager 1998).  A subclass of the A stars are
strongly magnetic -- the Ap and Am stars.  These represent $\sim
10$\% of the A-star class (MacGregor 2005), and are stars with
surface magnetic fields in the multi-kilo-Gauss range, often
modelled as oblique magnetic rotators in which the magnetic field is
predominantly dipolar (Stibbs 1950). The wind diagnostics,
particularly the portion trapped by the global dipolar magnetic
field loops, are likely to be found in the X-ray band, as the
trapped wind streams from opposite hemispheres are forced by the
magnetic field into supersonic collisions to generate shock-heated
gas (e.g., Babel \& Montmerle 1997; Townsend \& Owocki 2005). Again,
these types of stars are not often expected in finite-source
microlensing events.  The solar-type star observed in the extended
source event reported in Abe et al. (2003) presumably has a
solar-like stellar wind, with a feeble mass-loss rate of order
$10^{-14} \, M_\odot$ yr$^{-1}$. Consequently, wind spectral
features accessible to ground-based follow-up would be essentially
non-existent for such a source; instead, coronal diagnostics would
be found in the UV and X-ray bands.  Although one might expect to
see effects in Ca {\sc ii} H and K that are sensitive to
chromospheric emission, our model line profiles are not appropriate
for the quasi-static chromosphere.

\noindent The most likely class of stars to be found as sources in
microlensing transit events, either by a single or a binary lens,
are the red giant stars, a consequence of their being reasonably
common and fairly large in size.  Red giant winds have modest
mass-loss rates of order $10^{-10}$ or $10^{-8} M_\odot$ yr$^{-1}$;
however, even these estimates are not certain because the origin of
their winds is not well-understood. The absence of X-ray emission
for K and M giants (Linsky \& Haisch 1979) indicates that their
winds are not Parker-like, although they do show chromospheric
features.  The strong Mg {\sc ii} h and k lines and Fe {\sc ii}
lines can be used to probe cool star winds; however, these lines are
found in the UV which is not amenable to ground-based follow-up. It
may be that a combination of molecular opacity and some dust
formation drives the winds of red giants (Jorgensen \& Johnson
1992), and perhaps certain molecular lines could be used as wind
diagnostics.

\noindent Red supergiants (RSG) and asymptotic giant branch (AGB)
stars have more interesting wind features that could be studied in
microlensing transits.  These are stars with high mass-loss rates of
order $10^{-5} M_\odot$ yr$^{-1}$. They are radiatively driven, but
unlike the early-type stars that are line driven, these stars are
continuum driven because of substantial dust formation in their
dense and cool winds. Simmons et al. (2002) have explored the
possibility that scattering polarization could be used to probe
their winds, but of course this is a continuous opacity. Similar to
the red giants, UV lines can be used to probe the wind flow.  Some
of RSG and AGB stars also show maser emission lines at radio
frequencies (Elitzur 1992; Habing 1996; Lewis 1998). These are
thought to form primarily in shells. Consequently, the model lines
presented in this paper are less likely to be relevant, although the
considerations of Ignace \& Hendry (1999) might be more appropriate,
if suitably modified for the emissivity function of the maser.  Once
again, stars of the RSG and AGB classes are fairly rare.

\noindent There are two other classes of sources for which our line
profile models are more relevant.  One is the active galactic nuclei
(AGN) that produce strong wind flows (Arav et al. 1995; Proga,
Stone, \& Kallman 2000).  The (rest frame) UV spectra of quasars
show some quite strong P~Cygni profiles, such as in C {\sc iv},
suggesting wind speeds in excess of 10,000 km s$^{-1}$.  The flows
are likely line-driven similar to early-type stars.  Already, some
authors have modelled the characteristic emission line profile
effects that would be observed for different source models during
microlensing events (Popovi\'{c}, Mediavilla, \& Mu\~{n}oz 2001;
Abajas et al. 2002), ranging from Keplerian disks to relativistic
disks to simple shells and jets. The microlensing under
consideration was by a single deflector; however, our spherically
symmetric wind models are not directly applicable to these cases
that mostly involve non-spherical geometries.  Some have claimed the
detection of microlensing effects relevant to probing the AGN
accretion disk (e.g., Chae et al. 2001), although none have reported
effects pertinent to a wind component. We will not discuss this
class further, and have mentioned it only because the general
consideration of emission profile variations during microlensing
events involving wind flows has been considered in the context of
AGN.

\noindent The second class of objects are supernovae (and the likely
related gamma-ray bursters).  The explosions of stars spew gaseous
ejecta into space.  There is a phase in which the ejecta follows
homologous expansion, not because the flow is accelerating, but
because the ejecta is moving outward with a distribution of speeds.
Still, the flow can be modeled as a wind, and the spectra at
different phases shows numerous wind-features, ranging from strong
P~Cygni lines, to recombination lines, to forbidden lines
(Filippenko 1997).  Because supernovae (SNe) are so intrinsically
luminous at early times, they can be seen to large distances, and so
the probability of lensing increases owing to the considerable
cosmological path lengths involved.

\noindent Recently Dalal et al. (2003) have discussed the impact of
weak lensing due to large scale structure on the use of SNe as
standard candles. Other authors have considered microlensing as a
tool for studying the ejecta of SNe and even gamma-ray bursts
(GRBs). For example, Schneider \& Wagoner (1987) considered
polarimetric variations of pseudo-photospheres produced in SNe
during a microlensing event. Bagherpour, Branch, \& Kantowski (2005)
modelled the impact of microlensing on P Cygni line profiles in the
spectra of Type Ia SNe. Gaudi, Granot, \& Loeb (2001) have invoked
microlensing to explain a brightening event in the lightcurve of
GRB-000301C. However, thus far a more comprehensice study that
systematically explores the effects of microlensing on emission line
features formed in SN and GRB ejecta remains to be made; our
illustrative results presented here are a further step in that
direction.

\noindent With respect to SNe, there are several points of caution
to be made when applying the results of this paper. First, our
models assume a steady-state source.  Of course, the luminosities,
ejecta optical depth, and photospheric radii evolve with time in
real SNe. Our results are applicable to SNe if the lensing event is
relatively fast compared to photometric and structural changes of
the ejecta. Under what conditions will this be true?  Suppose we
take the duration of the lensing event to be one week. If the
fastest ejecta shell moves at $10^4$ km s$^{-1}$, then the extent of
the ejecta shell would be at least $r_{\rm max}\approx 40$~AU at
early phases of the explosion in which homologous expansion is
applicable. Moreover, the evolution must be such that the lens moves
across the ejecta shell (in projection) before the dimensions of the
photosphere change significantly. This means the proper motion of
the lens should greatly exceed that of the photospheric expansion.
Given the $10^4$ km s$^{-1}$ ejecta speed as a maximum, and assuming
a transverse velocity of $10^2$ km s$^{-1}$ for the lensing mass, we
require that $D_{\rm SN} \gg D_{\rm L}$. Consequently, if the lens
were located at kpc distances (e.g., in the halo, the bulge, or the
Large Magellanic Cloud), then the SN should be located at Mpc
distances (e.g. at the distance of the Virgo cluster).

\noindent As a consistency check, we compare the angular size of the
SN ejecta $\theta_{\rm SN}$ to $\theta_{\rm E}$, as given by

\begin{equation}
\frac{\theta_{\rm SN}}{\theta_{\rm E}} = \frac{r_{\rm max}/D_{\rm
SN}}
    {\sqrt{\left(R_{\rm L}/D_{\rm L}\right)\,\left(1-D_{\rm L}/
    D_{\rm SN}\right)}}.
\end{equation}

\noindent If we assume the lensing object to be typical of MACHOs of
about $0.5 M_\odot$, then the Schwarzschild radius of the lens
$R_{\rm L}$ will be of order 1 km.  If we further assume $r_{\rm
max} \approx 40$ AU, and $D_{\rm SN} \gg D_{\rm L}$, then

\begin{equation}
\frac{\theta_{\rm SN}}{\theta_{\rm E}} \approx \frac{100\;{\rm kpc}}
    {D_{\rm SN}},
\end{equation}

\noindent where $D_{\rm L} = 10$ kpc was assumed.  Thus, SNe at
distances of several to dozens of Mpc would have $\theta_{\rm SN}/
\theta_{\rm E} \approx 0.01 - 0.1$ at early phases in the SN when
homologous expansion remains valid. These happen to be values
typical of our model simulations.

\noindent A further assumption of our models has been the common
``core-halo'' treatment, in which we can approximate the photosphere
as a ``hard'' spherical boundary, and the wind features are all
formed exterior to this boundary.  Recent radiative transfer
simulation of Type II SNe by Dessart \& Hillier (2005)  show this
not to be true. Instead, the radiative transfer in SNe ejecta is
more akin to that of Wolf-Rayet stars, for which the continuum forms
in the wind flow.

\noindent The purpose of our models has not been to represent
rigorously any particular type of stellar wind or class of stars,
but to illustrate the generic effects that could be observed with
intensive follow-up programs for source transit microlensing events.
In so doing, models for two common types of lines have been
considered -- resonance P Cygni lines and recombination emission
lines, for both point lensing and caustic crossing events from
binary lensing. Whether the source is a hot star or a cool star,
whether the lines are in the X-ray band or the radio band, the
general properties of the line variations in terms of lead/lag times
relative to the photosphere, of equivalent width variations, and
changes in line profile shape are qualitatively all to be expected.
We imposed the homologous expansion as a representative flow
velocity law. Observationally, the velocity law is something that
would preferably be determined from the line variations as the
microlensing event evolved. With the photospheric crossing time
determined from the photometric variations, the variations of the
line equivalent width and profile changes with time could be
converted to radius, calibrated by the photospheric radius, to
deduce flow velocity and opacity variations.

\begin{acknowledgements}
R. Ignace gratefully acknowledges support for this work by a NSF
grant, AST-0354262. The authors gratefully acknowledge the helpful
comments of the anonymous referee.

\end{acknowledgements}

\end{document}